\documentclass[twocolumn,pre,10pt,amsmath,amssymb,nofootinbib,showpacs,superscriptaddress,floatfix]{revtex4}

\newcommand{\dfr}[2]{\frac {\displaystyle #1}{\displaystyle #2}}
\usepackage{graphicx}
\usepackage[hypertex]{hyperref}

\begin{document}
\title{Gauge theory approach to glass transitions}
\author{Mikhail Vasin}

\address{Physical-Technical Institute, Ural Branch of Russian Academy of Sciences, 426000 Izhevsk, Russia}
\begin{abstract}
This theory combines a thermodynamic approach with a dynamic one in order to describe glass transition. Glass transition is regarded as an inaccessible second order phase transition, which is interrupted because of premature critical slowing down, caused by the system's frustration. The frustration-induced vortices are present in the structure besides thermoactivated vortices, and prevent the development of the order parameter fluctuations, that leads to the critical slowing down the system kinetics at some temperature above the phase transition point.
\end{abstract}

\maketitle

\section{Introduction}

The formulation of the microscopic glass transition theory has remained one of the most intriguing but still unresolved problems of condensed matter physics for a long time. Many systems which manifest this phenomenon regardless of their nature enable us to conclude that this phenomenon does not depend on any microscopic details, but it is determined by the symmetry properties of the systems, as in the case of phase transitions.  On the other hand, this universality allows us to conclude that the glass transition is only the result of general dynamic properties of condensed matter. Therefore, the question ``Is the glass transition a phase transition, or is it the dynamic effect related to the limited diffusion dynamics?'' remains actual. One can concretize this problem by dividing it into some important questions that guide our presentation of glass-transition theories, which, for example, was formulated in \cite{Biroli}. In the present work we try to conciliate these positions, and answer some of the questions.

We assume, that there are two key conditions for the implementation of glass transition.
First of all, we believe, that the glass transition has common nature with the second order phase transition, which is spontaneous breaking continuous symmetry and starting an ordering process in the system structure. On the other hand, in contrast to the phase transition this ordering process stops because of frustration, which arises in this process. The presence of frustration is the second key condition for glass transition. The frustration gives rise to production of vortexes (spin vortexes in the spin systems, or disclinations in the undercooled liquids \cite{Riv, Nelson}) in the structure, which prevent the growth of the ordered regions.

The nonequilibrium dynamics of the slowing vortex system, which takes into account the interaction between these vortexes, can be described in terms of the gauge field theory. Therefore the old, and almost forgotten idea of the gauge field description of glass transition underlies in this work \cite{Riv, Volovik, Hertz}. Note, that this approach is very close ideologically with the ``frustrated-limited domain theory'' \cite{Kivelson, Tarjus}, but the former seems to be more convenient versus the latter one, because of the absence of long-range interaction, which is deleted from the theory at the expense of the gauge field introduction.

An important point of this theory is the presence of the vortexes in the structure. They can exist without frustration, but then their concentration tends to zero at the phase transition point, in which the relaxation time diverges.
We take into account, that additional, induced by the frustration, vortices are present in the structure besides the thermoactivated vortices. Therefore, close to the glass transition there is a finite nonzero density of vortices. We believe that above the glass transition temperature the vortices system is in equilibrium. This fact makes it possible to average over the gauge field sources, that results in the temperature displacement of the critical point in the gauge field subsystem. Then we use the renormalization group methods and critical dynamics methods for the description of the static and dynamic properties of the system at the new critical point.

In terms of this approach the glass transition corresponds to the critical point in the gauge field subsystem.
We examine both the static case and the dynamic case, and show that the theory's linear and non-linear  susceptibilities agree with the experimental data.
The dynamic features of this transition, such as the Vogel-Fulcher-Tamman relaxation time dependence on temperature, the characteristic temperature dependencies of the heat capacity, and the plateau in the time dependence of the order parameter correlation function, can be revealed by means of the functional methods of non-equilibrium dynamics.
In conclusion we consider the relationship of the suggested theory with the mode-coupling theory approach and the frustrated-limited domain theory, and discuss its physical interpretation.

\section{Model of supercooled liquid close to glass transition}

One of the most elegant methods of description of the frustrated systems is given by the gauge theory \cite{Volovik, Hertz, Riv, Kanazawa}.
The static action of the system close to the second order phase transition in the general case has the form of
\begin{equation*}
    S=\beta \int \left[\dfr 12(\partial_i {\bf s })^2+U({\bf s})\right]d{\bf r},
\end{equation*}
where $\bf s$ is some order parameter field, $\beta = 1/TK_B$, $K_B$ is the Boltzmann constant, and $U({\bf s})=\mu^2 {\bf s}^2 +v{\bf s}^4 $.
The structure of vitrescent systems is inhomogeneous both in relative orientation of local ordering elements and in density. Congruent connections between the orientations are introduced by means of the gauge field, $A^{a}_{\mu }$, when the ordinary derivative, $\partial_i{\bf s}$, is replaced by the covariant derivative, $D_i{\bf s}$, \cite{Hertz}:
\begin{equation*}
\displaystyle \partial_{i}s_a \to D_i s_a=\partial_{i}s_a+g\varepsilon_{abc} A_{bi}s_c,
\end{equation*}
where $g$ is the Frank index ($g\approx 0.2$ for a poly-tetrahedral packing), and $\varepsilon_{abc}$ is the Levi-Civita tensor. Besides, the system action should contain the gauge-invariant term $\sim (D_iA^{a}_{\mu })^2$. Therefore, one gets the gauge-symmetric action:
\begin{equation}\label{1}
    S=\beta \int\left[\dfr 12(D_i {\bf s })^2+U({\bf s})+\dfr 14F_{a\mu\nu}F_{a\mu\nu}\right]d{\bf r},
\end{equation}
where
\begin{equation*}
\displaystyle F_{a\mu\nu}=\partial_{\mu}A_{a\nu}-\partial_{\nu}A_{a\mu}+g\varepsilon_{abc} A_{b\mu}A_{c\nu}.
\end{equation*}
As it was noted above, this model was suggested in \cite{Volovik, Hertz, Riv} last century, but did not become widespread.

In more detail the example of the three-dimensional Heisenberg spin-glass model, whose order parameter, ${\bf s}$, is considered to be a local magnetization vector with the SO(3) continuous symmetry group, was considered in \cite{Vasin}.
With $\mu ^2>0$ the action of this model is invariant under the SO$(3)$ gauge transformations and $\langle{\bf s}\rangle =0$. However, with $\mu ^2<0$ the symmetry is explicitly broken, since the system can arbitrarily ``choose'' only one state from all equivalent states with the minimum energy $U({\bf s})$ potential situated on the $|{\bf s}|=i\mu/\sqrt{2v}$ sphere. We fix the vacuum by means of fixing a point on the sphere. The system is no longer symmetrical with respect to the SO$(3)$ gauge group, but it is invariant under the SO$(2)$ group of the rotation around some chosen axis.

\begin{figure}[h]
\label{fig0}\centering
   \includegraphics[scale=0.3]{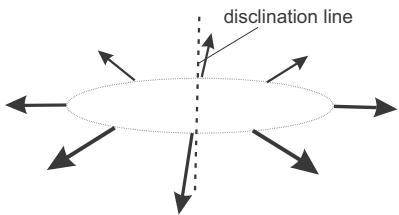}\\
   \caption{ Frustration induces topological defects (vortexes) in the system. These defects curve the space so that the local equilibrium configuration of spins contains the entire turn of their direction while going around this disclination along a closed contour. The availability of such vortexes in the system degenerates its collinear ground state. }\label{fig0}
\end{figure}

The expansion of the local magnetization field, ${\bf s}$, near one of the vacuum states, for instance $\left<{\bf s}\right>_0=(0,\,0,\,i\mu/\sqrt{2v})$, in small $\phi =s-i\mu/\sqrt{2v}$ deviations, and use of the gauge transformation properties allow to rewrite the action (\ref{1}) in the form of the functional of two massive vector bosons,  $A_{\kappa \mu }$ ($\kappa =\{1,\,2\}$), with the mass $M_0=ig\mu/\sqrt{2v}$, one massless vector boson, $A_{\mu }^3$, and one scalar field, $\phi $:
\begin{equation}
\begin{array}{c}
 \displaystyle S=\beta \int\left[\dfr 12(\partial_{\mu } \phi )^2+2\mu^2\phi^2+\dfr{g^2\mu^2}{4v}A_{\kappa\mu }A_{\kappa \mu }\right.\\[12pt]
 \displaystyle \left.+\dfr 14F_{a\mu\nu}F_{a\mu\nu}+v\phi^4+\dfr{g^2}2\phi^2A_{a\mu }A_{a\mu }\right]d{\bf r}.
\end{array} \label{RGE}
\end{equation}

One of the important features of the above model (\ref{RGE}) is the existence of soliton solutions, which correspond to the vortexes system (Fig.\,\ref{fig0}). These vortexes are the movable sources of the gauge field, and their system above the glass transition temperature can be considered as a system which remains in thermal equilibrium \cite{Riv}.
From (\ref{RGE}) one can see that the gauge field correlation function diverges in $T_c$ just as the scalar field correlation function, that corresponds to the sources concentration reaching zero at $T_c^{+}$.

The method of introduction of disorder into the theory is very important and plays a key part. In order for the system to possess glass properties we should inject disorder into this system, which
gives rise to the frustration of its structure. In approach \cite{Hertz} the disorder is associated directly with the quenched gauge field, and the $A_{a\mu }$ field is frozen in an arbitrary configuration with some $P(A_{a\mu })$ distribution function. It is supposed that the gauge field describes the frustrations. But it is not quite right either, since the presence
of the quenched gauge field in the system does not mean the presence of frustration yet.
The frustration leads to multifold degeneration of the structural states of the system. It produces the vortexes in the structure, which do not vanish in the low-temperature state. Therefore, one should introduce the frustration-induced vortexes into the model, in additional to the thermally activated vortexes.

The frustrated structure can be represented as vortexes passing through the frustration planes~\cite{Riv, Nelson, Nusinov} (see Fig.\,\ref{fig0}). These vortexes are the sources of the $A_{a\mu }$ gauge field, and should be injected in the model by means of some sources field, $J_{a\mu }$.
Therefore, in contrast to \cite{Hertz}, we believe that it is more correct to consider the source field, $J_{a\mu }$, but not the gauge field, $A_{a\mu }$, as the random field. In this case the $A_{a\mu }$ field remains dynamic. For illustration the $A_{a\mu }$ field can be interpreted as a local relative rotation of neighboring local ordered domains, which corresponds to their local equilibrium. The domain orientations can be movable, but their local equilibrium configuration around any frustration in the $\delta V$ volume, bounded by the $\delta S$ sphere, should satisfy the following condition:
\begin{equation}\label{Sourse}
    \dfr 12\oint\limits_{\delta S} F_{a\mu\nu}dS_{\nu }=\int\limits_{\delta V} J_{a\mu}d{\bf r}\neq 0,
\end{equation}
which follows from the principle of the least action. As a result one should add the term with the source field in the action:
\begin{equation}\label{41}
\begin{array}{c}
 \displaystyle S=\beta \int\left[\dfr 12(\partial_{\mu } \phi )^2+2\mu^2\phi^2+\dfr{g^2\mu^2}{4v}A_{\kappa\mu }A_{\kappa\mu }\right.\\[12pt]
 \displaystyle \left.+\dfr 14F_{a\mu\nu}F_{a\mu\nu}+v\phi^4+\dfr{g^2}2\phi^2A_{a\mu }A_{a\mu }+J_{a\mu } A_{a\mu }\right]d{\bf r}.
\end{array}
\end{equation}

Let us consider the model (\ref{RGE}) in the fluctuation region close to $T_c$.
Note, that this model was formulated for the low temperature state, $T<T_c$, when the system symmetry is already broken.
It is important, that in the high temperature state these reasonings are correct too. Mathematically it is formal extension of the theory to the high temperature region. 
One can understand the physical meaning if the critical fluctuations are presented in the dual representation as moving vortices in the ordered phase.
With approximation of $T$ to $T_c^{+}$ the density  of the mobile vortexes decreases.  The expression (\ref{RGE}) binds the critical fluctuation scale with the inter-vortex distance and vortex density, so that the effective linear size of the critical fluctuations diverges proportionally to the gauge field correlation length. However, as it will be shown below, the frustration changes this picture a bit.

At the temperatures above $T_c$ the structure is not a geometrical invariant, the vortexes (disclinations) can move, be born and be annihilated in it.
But the frustration leads to the existence of a non-zero minimal vortex density.  In terms of the field theory this means that any time in the system  there is finite density, $I_0$, of the gauge field sources. Therefore the system can not be frozen in an ordered state.
After freezing some vortices are quenched in the structure, but above this point the vortex subsystem is believed to be in thermal equilibrium \cite{Riv}.
Therefore, in the description of the state above the freezing point averaging over $J_{a\mu }$ leads to the redefinition of the partition function:
\begin{equation*}\label{SS}
    Z=\int \left[\int  \exp\left(-S-\dfr {\beta }4\int I_0^{-1}J_{a\mu }^2 d{\bf r}\right) \mathfrak{D}J_{a\mu }\right]\mathfrak{D}\phi \mathfrak{D}A_{a\mu },
\end{equation*}
where $\int \dots \mathfrak{D}x$ is the continual integral. It results in the additional contribution to the $A_{a\mu}$ ``mass'', which is as follows:
\begin{equation}\label{random2}
    M^2=M^2_0-I_0=\mu^2g^2/4v-I_0.
\end{equation}
Thus, the frustration leads to the renormalization of the gauge field mass.

The renormalization of the gauge field mass affects the critical behavior of the system, since it shifts $M^2=0$ singularity to the temperature range above the virtual phase transition point, $T_c$.
If we assume that $\mu^2=\alpha K_B(T-T_c)$, where $\alpha $ is some constant,  then from (\ref{random2}) we have the critical divergence of the $A_{\mu }^a$ field
correlation radius at $T_g=T_c+4I_0v/\alpha K_Bg^2$. In \cite{Vasin} it was shown that it results in the critical slowing-down of the fluctuations. Thus, the disorder-induced frustrations inhibit the growth of the $\phi$ field correlation length, and the system freezes in a disordered state.  The $T_g$ can be regarded as the glass transition temperature. We analize this supposition below.

\section{Analysis of the spin-glass model in static case}

As it was noted above, averaging over $J^a_{\mu }$ leads to shifting the transition temperature from $T_c$ to $T_g$.  In this case the mass of the gauge field becomes $M$, and the propagators of the theory fields are written in the following form:
\begin{gather*}
G_0(k)=\dfr{1}{k^2+\mu^2},\quad \Delta_{\mu\nu}=\dfr{\delta_{\mu\nu}+k_{\mu}k_{\nu}/k^2}{k^2+M^2}.
\end{gather*}
Let us choose the model parametrization so that $T_gK_B=1$ for simplicity.
Close to the glass transition $M\to 0$, and the gauge field correlator becomes the same as the propagator of a massless photon,
\begin{gather*}
\Delta_{\mu\nu}\simeq\dfr{\delta_{\mu\nu}}{k^2}.
\end{gather*}
It is known that the theory is renormalizable in this case. Therefore, one can investigate the critical properties of the model at $T\to {T_g}^+$.

The canonical dimensions of the values of our theory are given in the table:
\begin{center}
\begin{tabular}{|c|c|c|c|c|c|c|c|}
  \hline
  $F$ & $k$ & $\phi$ & $A$ & $\mu^2$ & $M^2$ & $g$ & $v$ \\ \hline
  $d_k[F]$ & 1 & $-1-\dfr {d_k}2$ & $-1-\dfr {d_k}2$ &  $d_k-2$ & $d_k-2$  & $0$ & $0$  \\ \hline
\end{tabular}
\end{center}
The renormalization procedure only refines these values, which results in the replacement of the canonical dimensions by the critical ones $\Delta [F]=d[F]+O[F](\varepsilon )$, where $\varepsilon = 4-d_k$, and $d_k$ is the space dimension.
\begin{figure}[h]
\centering
   \includegraphics[scale=0.6]{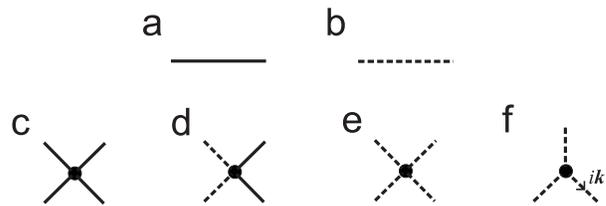}
   \label{f1}\\
   \caption{ The graph representation of the Green functions: a~---~$G_0$, b~---~$\Delta _{\mu\nu}$, and the action terms.}
\end{figure}



The renormalization procedure is carried out with the standard method \cite{Vas}.
It is assumed that the fields $\phi $ and $A$ are slow-varying ones, such that the Fourier-transformed fields have only long-wave components: $|k|<k_0$.
At the first step of RG transformations one integrates the partition function over the components of the fields in the limited wave band $\Lambda k_0<k<k_0$, where $\Lambda \ll 1$ is the regularization parameter (momentum cutoff). It is expected that under certain conditions this action has a structure similar to the original one, in this case a model is multiplicatively renormalizable. One can check that the formulated model satisfies this criterion. As a result one gets an effective action $S_{\Lambda }$ with renormalized parameters ($Z_{v}$, $Z_{\mu}$, $Z_{M}$, $Z_{g}$, and $Z_{g^2}$), which are named the ``constants of renormalization'' and depend on the cutoff $\Lambda $; at the second step, one makes the inverse scaling transformation of the fields ($Z_{A}$, $Z_{\phi }$) and coordinates, which aims at restoring the original cutoff scale $k_0$.
Then the renormalized parameters have the following form:
\begin{equation}\label{CH8R0}
\begin{array}{l}
   \displaystyle \mu^{2(R)}=Z_{\mu^2}Z^2_{\phi }\Lambda^{d}=\Lambda^{-2} Z_{\mu^2},\\
   \displaystyle M^{2(R)}=Z_{M^2}Z^2_{A }\Lambda^{d}=\Lambda^{-2 } Z_{M^2},\\
   \displaystyle g^{(R)}=Z_{g}Z_{A}^3\Lambda^{2d+1}=\Lambda^{-\varepsilon/2 } Z_{g},\\
   \displaystyle g^{2(R)}=Z_{g}^2Z_{A}^4\Lambda^{3d}=Z_{g}^2Z_{A}^2Z_{\phi}^2\Lambda^{3d+3z}
   =\Lambda^{-\varepsilon } Z_{g^2},\\
   \displaystyle v^{(R)}=Z_{v}Z_{\phi}^4\Lambda^{3d}=\Lambda^{-\varepsilon } Z_{v}.
\end{array}
\end{equation}
It is assumed that $S$ is invariant with respect to the above scale transformations at the critical point.

Close to $M^2=0$ ($T\gtrsim T_g$) the gauge field becomes massless, but the scalar field remains massive with $\mu^2=4I_0v/g^2$. Therefore, the contribution to the renormalization is made only by the loops of the gauge field propagators.

\begin{figure}[h!]
   \centering
   \includegraphics[scale=0.6]{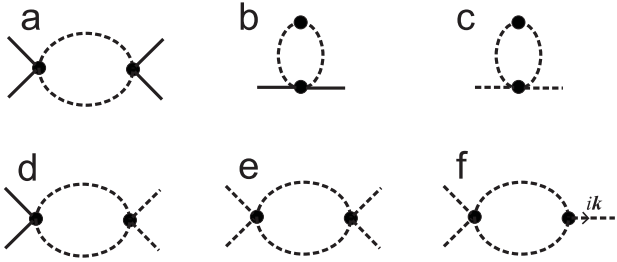}
   \caption{The graphical representation of one-loop contributions to the terms of the action.}
   \label{fig1S}
\end{figure}

In Fig.\,\ref{fig1S} some graphs giving logarithmically divergent contributions to the renormalized theory are  presented.

Let us explain in detail the renormalization of $M^2$ as an example. We will limit ourselves to using the one loop approximation, that is quite enough for the demonstration of all the features of the theory. In this case the renormalization constant of $M^2$ has the form
\begin{equation}\label{CH8R0}
\begin{array}{l}
  \displaystyle Z_{M^{2}}\delta_{\mu\nu}\simeq M^2\delta_{\mu\nu}-\dfr{6M^2g^2}{4(2\pi)^{d_k}}\int\limits_{\Lambda k_0}^{k_0} \Delta_{\mu\lambda}(k)\Delta_{\lambda\nu}(k)d{\bf k} \\[12pt]
  \displaystyle \simeq M^2\delta_{\mu\nu}-\dfr{3M^2g^2}{(2\pi)^{d_k}}\dfr{\delta_{\mu\nu}d_k\pi^{d_k/2}}{\Gamma(1+d_k/2)}\int\limits_{\Lambda k_0}^{k_0} \dfr{k^{d_k-1}}{k^4}dk.
\end{array}
\end{equation}
One can see that the integral in this expression introduces a logarithmically divergent contribution to $M^2$ renormalization if the momentum dimension is $d_k=4$. In this case we get following expression for the renormalized value of $M^2$:
\begin{equation}\label{CH8R}
M^{2(R)}=e^{2\xi} Z_{M^2}\simeq e^{2\xi}\left[ M^2-3\dfr{M^2g^2}{8\pi^2}\xi\right],
\end{equation}
where $\xi=\ln(1/\Lambda )$ is the logarithmically divergent factor.
In the same way one can get other terms of the renormalized action:
\begin{equation}\label{CH8R1}
\begin{array}{c}
   \displaystyle \mu^{2(R)}=e^{2\xi} Z_{\mu^2}\simeq e^{2\xi}\left[ \mu^2-\dfr{M^2g^2}{8\pi^2}\xi\right],\\
   \displaystyle g^{(R)}=e^{\varepsilon\xi/2 } Z_{g}\simeq e^{\varepsilon\xi/2 }\left[g-\dfr{g^4}{8\pi^2}\xi \right],\\
   \displaystyle g^{2(R)}=e^{\varepsilon\xi } Z_{g^2}\simeq e^{\varepsilon\xi }\left[g^2-\dfr{g^4}{4\pi^2}\xi \right],\\
   \displaystyle v^{(R)}=e^{\varepsilon\xi } Z_{v}\simeq e^{\varepsilon\xi }\left[v-\dfr{g^4}{8\pi^2}\xi \right],
\end{array}
\end{equation}
Hence, in the one-loop approximation the renormalization group of the model under study has the form:
\begin{equation}\label{RG1}
\begin{array}{l}
   \displaystyle \dfr{\partial \ln(M^2)}{\partial \xi}= 2-3g^2/8\pi^2,\\
   \displaystyle \dfr{\partial \ln(\mu^2)}{\partial \xi}=2-\dfr{M^2g^2}{8\mu^2\pi^2}\approx 2 ,\\
   \displaystyle \dfr{\partial \ln(g^2)}{\partial \xi}= \varepsilon -g^2/4\pi^2,\\
   \displaystyle \dfr{\partial \ln v}{\partial \xi}= \varepsilon -g^4/8v\pi^2.
\end{array}
\end{equation}
 From the condition of the stable point existence, ${\partial \ln (g^2)}/{\partial \xi}=0$, ${\partial \ln (v)}/{\partial \xi}=0$, we get
$g^2=4\pi^2\varepsilon$, and $v=g^2/2$. In addition, one can see that
\begin{equation}\label{F}
 M^2 = \dfr{\alpha K_Bg^2}{4v}(T-T_g)\approx e^{2\xi}.
\end{equation}
Thus, in $T_g$ the gauge field becomes massless, whereas the local magnetization field still has got the ``mass'', $\mu $. The analysis of the model close to this point allows us to assert that $T_g$ is
nothing else but the glass transition temperature. First of all, it is evident from (\ref{41}) that the linear susceptibility, $\chi_L =\partial \langle\phi\rangle/\partial h\sim \mu^{-2}=g^2/4I_0v$
($h$ is an external source of the field $\phi $), is finite at $T=T_g$.
Similar unsophisticated estimation gives the correlation length, $r_{cor}\sim \sqrt{g^2/4I_0v}$, is finite too.
One can see that in the absence of the frustration, $I_0=0$, the linear susceptibility and the correlation length diverge in $T_g$. This is natural, because in this case $T_g=T_c$ and the glass transition becomes the second order phase transition.

\begin{figure}[h]
   \centering
   \includegraphics[scale=0.5]{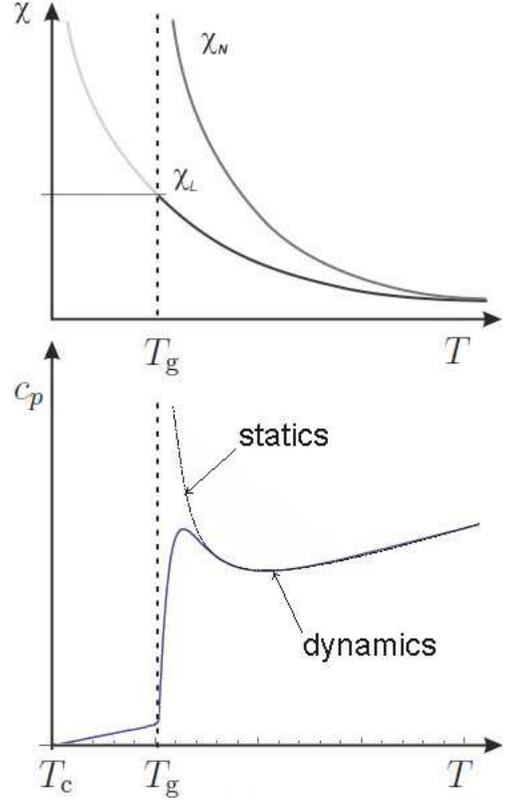}
   \caption{The qualitative presentation of the temperature function of the nonlinear susceptibility, $\chi_{N}\propto 1/(T-T_g)^{\gamma }$, and linear susceptibility, $\chi_L\propto 1/(T-T_c)$. The $\chi_L$ curve is not described by this theory at $T<T_g$, it is finite in this region.
   The qualitative picture of the temperature dependence of the heat capacity near $T_g$, which is obtained from the theory.}
   \label{capacity}
\end{figure}

On the other hand,
nonlinear susceptibility, $\chi_N=\partial^3\langle \phi\rangle/\partial h^3=\langle \phi^4\rangle$~\cite{Spinglasses}, diverges near $T_g$ because of the divergence of the
$A$-field loop contribution  (Fig.\,\ref{capacity}):
\begin{gather}\label{NSUS}
\chi_N\sim Z_{v}\sim e^{\varepsilon \xi}\sim (T-T_g)^{-\gamma },
\end{gather}
where $\gamma \approx \varepsilon/(2-3\varepsilon/2)$.
The estimation of this exponent in \cite{Vasin} was wrong because of incorrect summation of the graph series. The more accurate estimation gives $\gamma \approx 2$ when $\varepsilon =1$.

Using this theory allows us to determine the temperature dependence of the system heat capacity at $T\to T_g^+$:
\begin{equation}
 \begin{array}{c}
    \displaystyle c_p=\dfr{dU}{dT}\approx-K_B\ln Z-\dfr{K_BT}{Z}\dfr{dZ}{dT}= -K_B\ln Z\\[12pt]
    \displaystyle +\dfr{V\alpha g^2}{4v}
   \left(\dfr{\partial}{\partial T}-\dfr 1T\right)\left[(T-T_g)\langle A_{ij} A_{ij}  \rangle_{\bf r=0}\right]
    \\[12pt] \displaystyle
    +V\alpha\left(\dfr{\partial }{\partial T}-\dfr 1T\right)\left[(T-T_c)\langle \phi \phi  \rangle_{\bf r=0}\right] \\[12pt]
    \displaystyle +3V^2v\left(\dfr{\partial}{\partial T}-\dfr 1T\right)\left[ \langle \phi \phi \rangle_{\bf r=0}\langle \phi \phi \rangle_{\bf r=0}\right]\\[12pt]
    \displaystyle +V^2\dfr{g^2}{2}\left(\dfr{\partial}{\partial T}-\dfr 1T\right)\left[ \langle \phi \phi \rangle_{\bf r=0}\langle A_{ij}A_{ij}\rangle _{\bf r=0}\right]\\[12pt]
    \displaystyle +V^2\dfr{3g^2}{4}\left(\dfr{\partial}{\partial T}-\dfr 1T\right)\left[ \langle A_{ij}A_{ij} \rangle _{\bf r=0}\langle A_{ij}A_{ij}\rangle _{\bf r=0}\right].
\end{array}
\end{equation}
Since the matter keeps disordered structure at freezing, we will believe that $Z$ weakly depends on the temperature at $T_g$. Let us assume that the first term
is some constant. One can present the correlation functions in the form of
\begin{gather}
\langle A_{ij}A_{ij}\rangle_r = |r|^{-1} e^{-|r|M },\quad \langle \varphi\varphi \rangle_r = |r|^{-1} e^{-|r|m },
\end{gather}
where $M\propto (T-T_g)^{1/2}$, and $m\propto (T-T_c)^{1/2}$.
Therefore, close to $T_g$ the terms with $\dfr{\partial }{\partial T}\langle A_{ij}A_{ij}\rangle \propto (T-T_g)^{-1/2}$ are dominant, since they diverge at $T_g$, as opposed to the terms which contain only $\langle \varphi \varphi \rangle $.
As a result $c_p\propto (T-T_g)^{-1/2}$.

The above results show that $T_g$ is a critical point of the model, but they do not allow to conclude that it is the glass transition that occurs at $T_g$.  The correlation length of the $A$-field diverges in $T_g$, but the physical meaning of this correlation function is not clear. Besides, the obtained temperature dependence of the heat capacity near $T_g$ diverges and is not characteristic for glass transition.
One can show that this is a result of only the static character of the theory.
Below the analysis of the critical dynamics of this system close to $T_g$ allows to make sure in it, and to show that $T_g$ is the glass transition temperature.

\section{ Spin-glass model in dynamic case}

One can examine the non-equilibrium dynamics of the system close to $T_g$ by means of functional methods of non-equilibrium dynamics~\cite{Kamenev} near the critical point.  It leads to the representation of the partition function of the system in the form of
\begin{equation}\label{L2}
\displaystyle Z=\int \exp (-S^*)\mathfrak{D}\vec\phi \mathfrak{D}\vec A_{a\mu},
\end{equation}
with
\begin{equation}\label{L3}
\begin{array}{c}
\displaystyle
S^*=\frac 12\int \left[\vec\phi(t,\,{\bf r})\hat G^{-1}(t-t',\,{\bf r-r'})\vec\phi(t',\,{\bf r'})\right. \\[12pt]
   \displaystyle \left.  +\vec A_{a\mu }(t,\,{\bf r})\hat\Delta_{\mu\nu}^{-1}(t-t',\,{\bf r-r'})\vec A_{a\nu}(t',\,{\bf r'})\right]d{\bf r}d{\bf r'}dt dt' \\[12pt]
    +\displaystyle \int\left[ g\varepsilon_{abc}(\partial_{\mu}\bar A_{a\nu })A_{b\mu}A_{c\nu}  +g\varepsilon^{abc}(\partial_{\mu}A_{a\nu })\bar A_{b\mu}A_{c\nu}\right.\\[12pt]
   \displaystyle +g\varepsilon_{abc}(\partial_{\mu}A_{a\nu })A_{b\mu}\bar A_{c\nu}
   +g^2\varepsilon_{abc}\varepsilon_{aij}\bar A_{b\mu }A_{c\nu}A_{i\mu}A_{j\nu}
   \\[12pt]
   \displaystyle \left. +g^2 \bar A_{a\mu}A_{a\mu}\phi^2+g^2 (A_{a\mu})^2\bar\phi\phi +
    v4\, \bar\phi\phi^3\right]d{\bf r}dt,
\end{array}
\end{equation}
\begin{figure}
   \centering
   \includegraphics[scale=0.6]{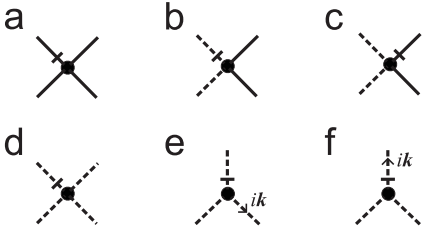}
   \caption{The graph representation of (\ref{L3}) action terms.}
   \label{fig2D}
\end{figure}
where $\vec\phi=\left\{ \bar\phi ,\,\phi  \right\}$, and $\vec A_{a\mu}=\left\{ \bar A_{a\mu},\,A_{a\mu} \right\}$ are vectors, the components of which are named ``quantum'' and ``classical'',
respectively, in the Keldysh representation~\cite{Kamenev}, $G^{-1}$ and $\Delta_{\mu\nu}^{-1}$ are matrices, inverse to the Green functions matrices having the following form:
\begin{multline}\label{eq:G0-1}
    \hat G=\left(\begin{array}{rl}
                            G^K_0 & G^A_0 \\
                            G^R_0 & 0
                          \end{array}\right), \quad
    \hat \Delta_{\mu\nu}=\left(\begin{array}{rl}
                            \Delta ^K_{\mu\nu} & \Delta ^A_{\mu\nu} \\
                            \Delta ^R_{\mu\nu} & 0
                          \end{array}\right),
\end{multline}
where
\begin{multline}\label{7}
   \displaystyle G^{R(A)}_0 (k,\,\omega )
   =\dfr{1}{k^2+\mu^2\pm i\Gamma_{\phi}\omega },\\
   \displaystyle G^{K}_0 (k,\,\omega )=\dfr{2\Gamma_{\phi}}{(k^2+\mu^2)^2+\Gamma_{\phi}^2\omega^2 },
\end{multline}
$\Gamma_{\phi} $ is the kinetic coefficient of the local magnetization.
In the case of $M\to 0$
\begin{multline}\label{8}
    \displaystyle \Delta^{R(A)} _{\mu\nu}(k,\,\omega )\simeq\dfr{\delta_{\mu\nu}}{k^2\pm i\Gamma_{A}\omega },\\
    \displaystyle \Delta^{K} _{\mu\nu}(k,\,\omega )\simeq\dfr{2\Gamma_{A} \delta_{\mu\nu}}{k^4+\Gamma_{A}^2\omega^2 },
\end{multline}
where $\Gamma_A $ is the kinetic coefficient of the gauge field.
\begin{figure}[h!]
   \centering
   \includegraphics[scale=0.6]{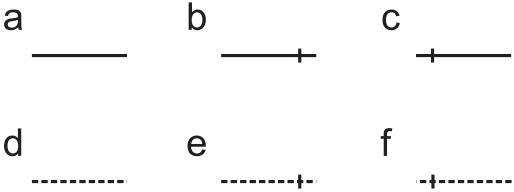}
   \caption{The graph representation of the Green functions: a~---~$G^K_0$, b~---~$G^A_0$, c~---~$G^R_0$, d~---~$\Delta ^K_{\mu\nu}$, e~---~$\Delta ^A_{\mu\nu}$, f~---~$\Delta ^R_{\mu\nu}$.}
   \label{fig0D}
\end{figure}

\section{Critical dynamics of the model}

The critical behavior of the system close to the $T_g$ can be considered within the critical dynamics technique~\cite{Vasin2, Vasin}.
The critical dynamics rests on the hypothesis of dynamic scaling, according to which the action should be invariant under scale transformations, which conformly expand the space and time coordinates ($\omega \sim k^{d_{\omega }}$). In this case the summarized dimension, $d=d_k+d_{\omega }$ ($d_{\omega}=z$ is the dynamic exponent), has the same role as the conventional (momentum) dimension, $d_k$, in the static case. The canonical dimensions of the fields and the model parameters are determined from the condition of dimensionless action. The corresponding summarized canonical dimensions, $d[F]$, of any values, $F$, depend as
$$
d[F]=d_k[F]+z\cdot d_{\omega }[F],
$$
where $d_{\omega }[F]$ is the frequency dimension \cite{Vas,Pat}.
The canonical dimensions of the values of our theory are given in the table:


\begin{center}
\begin{tabular}{|c|c|c|c|c|c|c|}
  \hline
  $F$ & $k$ & $\omega$ & $\phi$ & $\bar{\phi}$ & $A$ & $\bar A$  \\ \hline
  $d_k[F]$ & 1 & 0 & $-1-\dfr {d_k}2$ & $-1-\dfr {d_k}2$ & $-1-\dfr {d_k}2$ & $-1-\dfr {d_k}2$  \\ \hline
  $d_{\omega }[F]$ & $0$ & $1$ & $-1$ & $0$ & $-1$ & $0$  \\ \hline
  $d[F]$ & $1$ & $z=2$ & $-3-\dfr {d_k}2$ & $-1-\dfr {d_k}2$ & $-3-\dfr {d_k}2$ & $-1-\dfr {d_k}2$ \\
  \hline
\end{tabular}

\begin{tabular}{|c|c|c|c|c|c|c|}
  \hline
  $F$ & $g$ & $v$ & $\Gamma_A$ & $\Gamma_{\phi}$ & $\mu^2$ & $M^2$ \\ \hline
  $d_k[F]$ & $0$ & $0$ & $d_k-2$ & $d_k-2$ & $d_k-2$ & $d_k-2$ \\ \hline
  $d_{\omega }[F]$ & $0$ & $0$ & $-1$ & $-1$ & $0$ & $0$ \\ \hline
  $d[F]$ & $0$ & $0$ & $d_k-4$  & $d_k-4$ & $d_k-2$ & $d_k-2$ \\
  \hline
\end{tabular}
\end{center}
The renormalization procedure only refines these values, that leads to the replacement of the canonical dimensions by the critical ones $\Delta [F]=d[F]+O[F](\varepsilon )$.

The renormalization procedure is carried out with the standard method.
It is assumed that the fields $\phi $, $\bar\phi $, $A$, and $\bar A$ are slow-varying ones, such that the Fourier-transformed fields have only long-wave components: $|k|<k_0$; $\omega <\omega_0$.
At the first step of RG transformations one integrates the partition function over the components of the fields in the limited wave band $\Lambda k_0<k<k_0$, $\Lambda^z \omega_0<\omega <\omega _0$. The renormalized parameters have the following form:
\begin{equation}\label{CH8R0}
\begin{array}{l}
   \displaystyle \Gamma_{\phi}^{(R)}=Z_{\Gamma_{\phi }}Z_{\bar{\phi }}Z_{\bar{\phi }}\Lambda^{d+z}=\Lambda^{d+2-2(1+d/2)}Z_{\Gamma_{\phi }},\\
   \displaystyle \mu^{2(R)}=Z_{\mu^2}Z_{\bar{\phi }}Z_{\phi }\Lambda^{d+z}=\Lambda^{-2} Z_{\mu^2},\\
   \displaystyle M^{2(R)}=Z_{M^2}Z_{\bar{A}}Z_{A }\Lambda^{d+z}=\Lambda^{-2 } Z_{M^2},\\
   \displaystyle \Gamma_A^{(R)}=Z_{\Gamma_{A}}Z_{\bar{A}}Z_{\bar{A}}\Lambda^{d+z}=\Lambda^{\varepsilon } Z_{\Gamma_{A}},\\
   \displaystyle g^{(R)}=Z_{g}Z_{\bar{A}}Z_{A}^2\Lambda^{2d+2z+1}=\Lambda^{-\varepsilon/2 } Z_{g},\\
   \displaystyle g^{2(R)}=Z_{g}^2Z_{\bar{A}}Z_{A}^3\Lambda^{3d+3z}
   =\Lambda^{-\varepsilon } Z_{g^2},\\
   \displaystyle g^{2(R)}=Z_{g}^2Z_{\bar{A}}Z_{A}Z_{\phi}^2\Lambda^{3d+3z}
   =\Lambda^{-\varepsilon } Z_{g^2},\\
   \displaystyle v^{(R)}=Z_{v}Z_{\bar{\phi }}Z_{\phi}^3\Lambda^{3d+3z}=\Lambda^{-\varepsilon } Z_{v}.
\end{array}
\end{equation}

As well as in the static case, close to $M^2=0$ ($T\simeq T_g$) the gauge field becomes massless, but the scalar field remains the massive with $\mu^2=4I_0v/g^2$. Therefore, the contribution to the renormalization is made only by the loops of the gauge field propagators.
According to the separation of massive field theorem \cite{Collins} the Feynman diagrams, containing the propagators of the field, the mass of which is appreciably larger than the external momentum, are inversely proportional to the degree of this mass, and make a finite contribution to the renormalization.
In Fig.\,\ref{ff} some graphs giving logarithmically divergent contributions to the renormalized theory are  presented.

In the one-loop approximation the renormalization constant of $M^2$ has the form
\begin{multline}\label{CH8R1}
  \displaystyle Z_{M^{2}}\delta_{\mu\nu}\simeq M^2\delta_{\mu\nu}\\ -\dfr{6M^2g^2}{(2\pi)^{d_k+1}}\int\limits_{\Lambda k_0}^{k_0} \Delta^R_{\mu\lambda}(k,\,\omega)\Delta^K_{\lambda\nu}(k,\,\omega)d{\bf k}d\omega \\
  \displaystyle \simeq M^2\delta_{\mu\nu}-\dfr{12M^2g^2}{(2\pi)^{d_k+1}}\int\limits_{\Lambda k_0}^{k_0} \dfr{\dfr{d_k\pi^{d_k/2}}{\Gamma(1+d_k/2)}\delta_{\mu\nu}k^{d_k-1}}{(k^4+\omega^2 )(k^2+ i\omega )}dkd\omega \\[12pt]
  \displaystyle =M^2\delta_{\mu\nu}-\dfr{3M^2g^2}{(2\pi)^{d_k}}\dfr{\delta_{\mu\nu}d_k\pi^{d_k/2}}{\Gamma(1+d_k/2)}\int\limits_{\Lambda k_0}^{k_0} \dfr{k^{d_k-1}}{k^4}dk.
\end{multline}
The integral in this expression introduces a logarithmically divergent contribution to $M^2$. In this case the expression for the renormalized value of the $M^2$ coincides with (\ref{CH8R}):
\begin{figure}[h]
   \centering
   \includegraphics[scale=0.6]{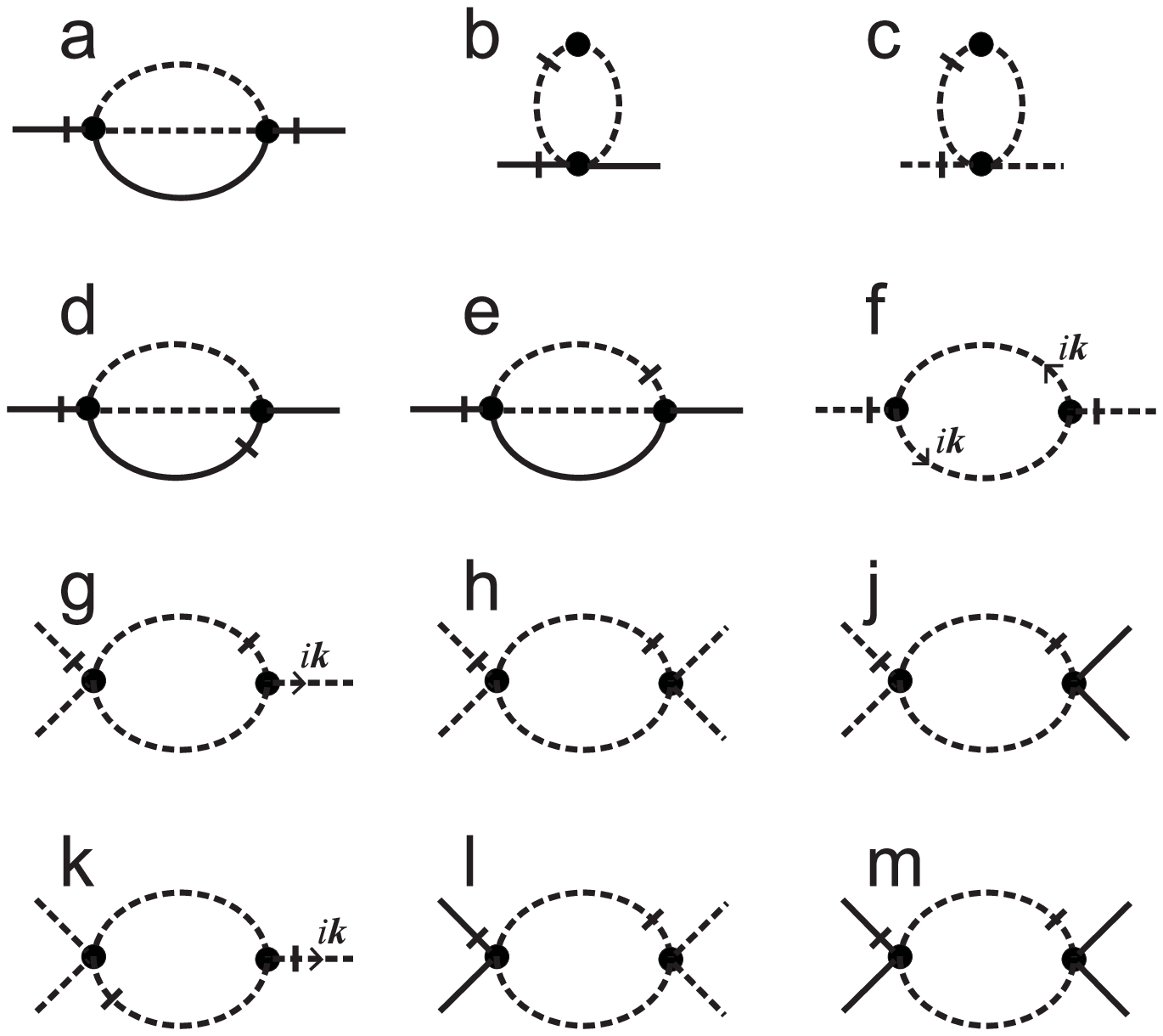}
   \caption{The graph representation of the one-loop contributions to the terms of the action.}
   \label{ff}
\end{figure}

In the same way one can get other terms of the renormalized action:
\begin{equation*}
\begin{array}{l}
   \displaystyle \mu^{2(R)}=e^{2\xi} Z_{\mu^2}\simeq e^{2\xi}\left[ \mu^2-\dfr{M^2g^2}{8\pi^2}\xi\right],\\
   \displaystyle \Gamma_A^{(R)}=e^{-\varepsilon\xi } Z_{\Gamma_{A}}\simeq e^{-\varepsilon\xi } \left[\Gamma_A+3\dfr{\Gamma_Ag^4}{4\pi^2}\xi +\dfr{\Gamma_Ag^2}{8\pi^2}\xi \right],\\
   \displaystyle g^{(R)}=e^{\varepsilon\xi/2 } Z_{g}\simeq e^{\varepsilon\xi/2 }\left[g-\dfr{g^4}{8\pi^2}\xi \right],\\
   \displaystyle g^{2(R)}=e^{\varepsilon\xi } Z_{g^2}\simeq e^{\varepsilon\xi }\left[g^2-\dfr{g^4}{4\pi^2}\xi \right],\\
   \displaystyle v^{(R)}=e^{\varepsilon\xi } Z_{v}\simeq e^{\varepsilon\xi }\left[v-\dfr{g^4}{8\pi^2}\xi \right],
\end{array}
\end{equation*}
where $\varepsilon = 4-d_k$.

One can see that the considered theory is renormalizable, since the renormalization procedure leads to the correction of only existing terms of the Lagrangian, and all renormalized parameters logarithmically diverge when $d_k=4$, which thereby is the critical dimension.

\begin{figure}[h]
   \centering
   \includegraphics[scale=0.6]{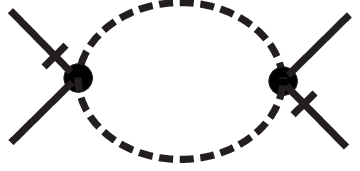}
   \caption{}
   \label{fig1}
\end{figure}

The interaction of the local magnetization fluctuations per gauge field (Fig.\,\ref{fig1}) plays the key part in the considered theory.
This becomes clear when we consider in detail the contribution to the renormalization of the $\Gamma_{\phi}$ node of the $a$ diagram, which is shown in Fig.\,\ref{ff}. This term is also interesting because the $\Gamma_{\phi}$ node is proportional to the relaxation time of the local magnetization field, and determines the kinetics of the glass-transition: the loop of the light field, $A^a_{\mu }$, which is given in Fig.\,\ref{fig1}, makes the logarithmically divergent contribution $\sim \ln(1/\Lambda )\delta(\omega )$. This term determines the divergence contribution of the $a$ graph (Fig.\,\ref{ff}) in which the massive field loop leads only to multiplying the logarithm by $4I_0v/g^2$ factor:
\begin{equation}\label{9}
\begin{array}{c}
    \displaystyle Z_{\Gamma_{\phi}} \approx \Gamma_{\phi}+ \dfr{4g^2I_0v\ln (1/\Lambda)}{\pi^2}\int\limits_0^{t_o}e^{-4I_0v|t|/\Gamma_{\phi} g^2}dt\\[12pt]
    \displaystyle =\Gamma_{\phi} +\dfr{g^4 \ln (1/\Lambda)}{\pi^2}(1-e^{-4I_0v|t_0|/\Gamma_{\phi} g^2}),
\end{array}
\end{equation}
where $t_o$ is the time of the observation of the system.
One can see that for $I_0\to 0$, or with a short observation time, $\Gamma_{\phi} g^2/4I_0v\gg t_o$, this contribution becomes negligibly small, and the theory becomes nonrenormalizable. This relates to the symmetry properties of the Yang-Mills model. In this case the fluctuation dissipation theorem (FDT) is always broken because of the free energy transfer between the local magnetization modes and massless gauge field modes (Goldstone modes).
However, there is also some problem in the presence of the quenched disorder: this node becomes nonlocal in time. This formally makes the renormalization group equations nonautonomous for $t_o\approx \Gamma_{\phi} g^2/4I_0v$, which violates the correctness of the renormalization procedure. The nonautonomous renormalization group has recently been considered in physical problems rather often (see, e.g., \cite{Beloborodov,Burm}). In order to avoid this problem we note that the renormalization group equations are autonomous in two limit cases: $t_o\ll \Gamma_{\phi} g^2/4I_0v$ and $t_o\gg \Gamma_{\phi} g^2/4I_0v$, the physical solution is the result of matching these two limit cases.

At large time scales, $t_o\gg \Gamma_{\phi} g^2/4I_0v$, the contribution to renormalization of $\Gamma_{\phi }$  is logarithmically divergent. Hence, in the one-loop approximation the renormalization group of the model under study has the form:
\begin{equation}\label{RG1}
\begin{array}{l}
   \displaystyle \dfr{\partial \ln(\Gamma_{\phi} )}{\partial \xi}={g^4 }/{\pi^2}, \\
   \displaystyle \dfr{\partial \ln(\Gamma_A )}{\partial \xi}=-\varepsilon +3g^4/4\pi^2 +g^2/8\pi^2 ,\\
   \displaystyle \dfr{\partial \ln(M^2)}{\partial \xi}= 2-3g^2/8\pi^2,\\
   \displaystyle \dfr{\partial \ln(\mu^2)}{\partial \xi}=2-\dfr{M^2g^2}{8\mu^2\pi^2}\approx 2 ,\\
   \displaystyle \dfr{\partial \ln(g^2)}{\partial \xi}= \varepsilon -g^2/4\pi^2,\\
   \displaystyle \dfr{\partial \ln v}{\partial \xi}= \varepsilon -g^4/8v\pi^2.
\end{array}
\end{equation}
As well as in the static case, from the condition of the stable point existence, ${\partial \ln (g^2)}/{\partial \xi}=0$, ${\partial \ln (v)}/{\partial \xi}=0$, we obtain
$g^2=4\pi^2\varepsilon$, and $v=g^2/2$.

At small time scales, $t_o\ll \Gamma_{\phi} g^2/4I_0v$, the contribution to the $\Gamma_{\phi }$ renormalization is negligibly small. In this case the RG-equations form similarly to (\ref{RG1}) except the first equation, which has the form:
\begin{equation}
\label{RG2}
\dfr{\partial \ln(\Gamma_{\phi} )}{\partial \xi}= 0.
\end{equation}

In order to match the solutions found in (\ref{RG1}) and (\ref{RG2}) it is necessary to analyze the renormalization in the $t_o\sim \Gamma_{\phi} g^2/4I_0v$ region, and get a matching function, $\Phi(\xi)$, as it was done in \cite{Beloborodov,Vasin}:
\begin{equation*}
\dfr{\partial \ln(\Gamma_{\phi} )}{\partial \xi}={g^4 }\Phi(\xi)/{\pi^2},
\end{equation*}
where $\xi\sim 1$. The matching function, which is $1$ in the case of large time scales, $\xi\gg 1 $, and $0$ in the case of small time scales, $\xi \ll 1$, can be chosen in the form:
\begin{equation*}
\Phi(\xi )=1-\Lambda^z=1-\exp(-z\xi),
\end{equation*}
where $z\approx 2$ is the dynamical index \cite{Vas,Pat}. As a result
\begin{equation}\label{VFT}
    \tau_{rel}=\Gamma_{\phi} \propto\exp\left(\dfr{2vg^2T_g}{\alpha \pi^2(T-T_g)}\right).
\end{equation}
Hence, critical slowing down of all relaxation processes does occur at $T_g$, and follows the Vogel-Fulcher-Tammann relation, which in our case was derived from the microscopic reasons by means of the Keldysh technique and critical dynamics method.

Note that if frustrations are absent, $I_0\to 0$, the freezing temperature coincides with the phase transition temperature. Then the diagrams with the loops of $\phi$ and $\bar\phi$ fields become divergent, and the system experiences the paramagnetic--ferromagnetic phase transition, which is described within the standard critical dynamics~\cite{CritDyn}.

\begin{figure}[h!]
   \centering
   \includegraphics[scale=0.35]{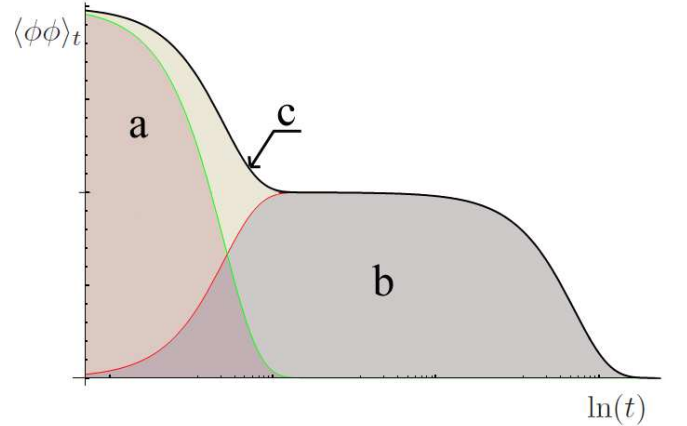}
   \caption{The dependence $\langle \phi \phi\rangle_t$ on $\ln (t)$: a) is the contribution of $\langle \phi \phi\rangle_t^0$, which is the
   Debye-relaxation; b) is the contribution of the second term which is given by the cooperative effects; c) is the sum of the first and second terms.}
\label{BosonPeak}
\end{figure}

In the one-loop approximation the $\phi$-field correlation function can be represented as the sum of the unperturbed and cooperative parts:
\begin{multline}\label{BP}
\displaystyle \langle \phi \phi\rangle_{t,\,k=0} =G^K(t) \simeq G^K_0(t)+\\ \dfr{2I_0g^2v\ln(L/a_0)}{\Gamma_{\phi}\pi^2}\int\limits_0^{t} G^K_0(t')e^{-4I_0v|t-t'|/\Gamma_{\phi} g^2} dt',
\end{multline}
where  $L$ is the size of the system, and $a_0$ is the interatomic distance.
This function has the form which is characteristic for glass systems (Fig.\,\ref{BosonPeak}).

The approximation of the linear and nonlinear susceptibilities in the dynamic theory coincides with that the static one. However, the dynamic theory allows us to improve the temperature dependence of the system heat capacity at $T\to T_g^+$. Carrying out the analysis, it is convenient to consider the correlation functions in the ($t,\,k$)-representation, and take into account that
$\langle \bar\phi \bar\phi  \rangle=\langle \bar A_{ij}\bar A_{ij}  \rangle=0$. As a result we obtain:
\begin{equation*}
 \begin{array}{c}
    \displaystyle c_p=\dfr{dU}{dT}\approx-K_B\ln Z-\dfr{K_BT}{Z}\dfr{dZ}{dT}= -K_B\ln Z\\[12pt]
    \displaystyle +\dfr{V\alpha g^2}{4v}
   \left(\dfr{\partial}{\partial T}-\dfr 1T\right)\left[(T-T_g)\int_k\langle \bar A_{ij}A_{ij}  \rangle_{t=0}\right]
    \\[12pt] \displaystyle
    +V\alpha\left(\dfr{\partial}{\partial T}-\dfr 1T\right)\left[(T-T_c)\int_k\langle \bar \phi \phi  \rangle_{t=0}\right] \\[12pt]
    \displaystyle +V\left(\dfr{\partial}{\partial T}-\dfr 1T\right)\left[ g^2\int_k\langle \bar \phi \phi \rangle_{t=0}\int_k\langle A_{ij}A_{ij}\rangle_{t=0}\right.
    \\[12pt]
    \displaystyle \left.  +g^2\int_k\langle \phi \phi\rangle_{t=0}\int_k\langle  \bar A_{ij}A_{ij}\rangle_{t=0}+ 12v\int_k\langle \bar\phi \phi\rangle_{t=0}\int_k\langle \phi\phi\rangle_{t=0}\right. \\[12pt]
     \displaystyle \left.+ 3g^2\int_k\langle \bar A_{ij}A_{ij}\rangle_{t=0}\int_k\langle A_{ij}A_{ij}\rangle_{t=0}\right],
\end{array}
\end{equation*}
where the $\int_k(\dots)=\int_{-\infty}^{\infty}(\dots)d{\bf k}$ notation is introduced.
Since the matter keeps disordered structure at freezing, we will believe that $Z$ weakly depends on the temperature at $T_g$. Therefore, let us assume that the first term
is some constant.  From (\ref{7}) and (\ref{8}) it is not difficult to get:
\begin{multline}
\displaystyle \langle \bar\phi\phi\rangle_t =\theta(t)\Gamma^{-1}_{\varphi}(T)e^{-t\varepsilon_k(m)/\Gamma_{\varphi}(T)},\\
\displaystyle \langle \phi\phi\rangle_t =\dfr {1}{\varepsilon_k(m)}e^{-|t|\varepsilon_k(m)/\Gamma_{\varphi}(T)},
\end{multline}
\begin{multline}
\displaystyle \langle \bar A_{ij}A_{ij}\rangle_t=\theta(t)\Gamma^{-1}_A(T)e^{-t\varepsilon_k(M)/\Gamma_A(T)},\\
\displaystyle \langle A_{ij}A_{ij}\rangle_t=\dfr {1}{\varepsilon_k(M)}e^{-|t|\varepsilon_k(M)/\Gamma_A(T)},
\end{multline}
where $\varepsilon _k(m)=k^2+m^2$,
whence
$$
\displaystyle\int_k\langle \bar\phi\phi\rangle_{t=0} =\dfr 1{V\Gamma_{\varphi}(T)}, \quad \int_k\langle  \bar A_{ij}A_{ij}\rangle_{t=0} =\dfr 1{V\Gamma_A(T)}.
$$
Consequently, we obtain:
\begin{multline*}
    \displaystyle c_p= -K_B\ln Z+\dfr{\alpha g^2}{4v\Gamma_A}
    +\alpha\left(\dfr{\partial}{\partial T}-\dfr 1T\right)\left[\dfr{(T-T_c)}{\Gamma_{\varphi}(T)}\right] \\
    \displaystyle +\left(\dfr{\partial}{\partial T}-\dfr 1T\right)\left[ \dfr {g^2}{\Gamma_{\varphi }(T)}\int_k \varepsilon_k^{-1}(M)+
   \dfr {g^2}{\Gamma_{A}(T)}\int_k\varepsilon_k^{-1}(m)\right. \\+\left. \dfr {3g^2}{\Gamma_{A}(T)}\int_k\varepsilon_k^{-1}(M)+\dfr {12v}{\Gamma_{\varphi}(T)}\int_k\varepsilon_k^{-1}(m)\right].
\end{multline*}
Since the system is close to $T_g$, then the last term gives the most considerable contribution in the heat capacity. One can obtain the qualitative form of the temperature dependence of the heat capacity close to the glass transition:
\begin{multline*}
    \displaystyle c_p (T)\propto \dfr{\partial }{\partial T}\left[e^{-CT_g/(T-T_g)}\right]
    =\dfr {e^{-CT_g/(T-T_g)}}{(T-T_g)^2}CT_g,
\end{multline*}
where $C=2\upsilon g^2/\alpha \pi^2$.
The graph of this expression is shown in Fig.\,\ref{capacity}.
In this picture one can see the sharp growth of the heat capacity near the glass transition which then drops at $T_g$, which is characteristic for vitrescent systems.

According to the suggested theory one can give a simple qualitative explanation of this dependence: when the system temperature approaches at $T_c$, the scale of the thermal fluctuations increases, and consequently, the heat capacity increases too, as in the case of the second order phase transition.
However, critical slowing down inhibits the growth of these fluctuations. The maximal size of the them is determined by the frustration scales. As a result the heat capacity does not infinitely diverge at $T_g$, but gets a finite maximum at $T_{\mbox{max}}>T_g$, and then falls down (Fig.\,\ref{capacity}).

\section{Derivation of the Mode-Coupling Theory equation}

As  the suggested gauge theory is based on the microscopic conceptions and has a general character, so it is natural to trace its relation with other known theoretical approaches to the description of glass transition.
The most applicable of them is the Mode-Coupling Theory, which, in the Zwanzig-Mori representation has the form of the motion equation for the pair correlation function.
The key problem of this theory is the determination of the memory function, which is the kernel in the integral-differential equation.

In order to derive the motion equation for the model structural factor, $S(t)\equiv G_K(t)$, in the presented theory, one can use the Dyson equations for $G^A$ and $G^R$:
\begin{gather}
G^A=G^{A(R)}_0+G^{A(R)}_0\otimes D^{A(R)}\otimes G^{A(R)},
\end{gather}
where $\otimes$ denotes the convolution of two functions on $t$, and
\begin{gather}
D^{A(R)}(t)=\theta(\pm t)\dfr{g^4\mu^2\ln(L/a_0)}{2\Gamma_{\phi}\pi^2}e^{\mp4I_0vt/\Gamma_{\phi} g^2}
\end{gather}
is the self-energy part, which has the physical meaning of the memory function in the mode coupling theory.
After the $\hat G^{-1}_0=\Gamma_{\phi}\partial_t-\nabla^2+\mu^2$ operator action on these functions one can subtract the first equation from the second one:
\begin{multline}
G^{-1}_0(G^R-G^A)=D^R\otimes G^R-D^A\otimes G^A=\\
=\int\limits_0^tD^R(|t'|)G^R(|t-t'|)dt'=\\=-\int\limits_0^tD^R(|t'|)\partial_tG^K(t-t')dt'.
\end{multline}
Then, using the FDT, one can obtain:
\begin{gather*}
\partial_t^2 G^K(t)+\mu^2\partial_tG^K(t)+T\int\limits_0^tD^R(|t'|)\partial_tG^K(t-t')dt'=0.
\end{gather*}
In order to present this expression in the typical form, one should add the static part to the Green function and memory function, $D(t)\to D(t)+\Omega^2/T$.
$\Omega $ is the microscopic frequency, obtained from the experimental static structural factor. Then we have:
\begin{multline*}
\partial_t^2 G^K(t)+\Omega^2G^K(t)+\mu^2\partial_tG^K(t)\\+T\int\limits_0^tD^R(|t'|)\partial_tG^K(t-t')dt'=0.
\end{multline*}
This equation being exact, the crux of the mode-coupling approach consists in formulating an approximate expression for $D^R(|t'|)$. In the considered theory this function can be approximated by the sum of the diagrams which give contribution to the self-energy part of the Green function.

\section{relation with the frustrated-limited domain theory}

The suggested theory is related very closely both with the Stillinger's ``tear and repair'' mechanism for relaxation~\cite{Stillinger} and the ``frustrated-limited domain theory'' of Kivelson and Tarjus \cite{Kivelson}. In these theories the frustration is described as the source of the strain free energy that opposes the spatial extension of the locally preferred structure and grows with the system size.  It results in breaking up the liquid structure into domains, whose sizes are limited by frustration with decreasing temperature.

One can switch over from the model (\ref{41}) to the frustrated-limited domain theory. Let us not average over $J$-field, as it was done above, and suppose that the disclinations are arbitrarily located in the structure. Instead of this let us get rid of the gauge field and carry out the integration of (\ref{41}) over $A_{\mu }^a$.
In order to simplify the consideration one can make a rather strong approximation, that the system relaxation depends only on the kinetics of the vortex system, and one can neglect the non-quadratic terms in (\ref{41}). Then the functional integration over $A_{\mu }^a$,
\begin{gather*}
Z_{A}=\int \mathfrak{D}A_{\mu }^a \exp\left[ -\int \left(\dfr 12A_{\mu }^a\Delta ^{-1}_{\mu\nu}A_{\nu }^a+J_{\mu }^aA_{\mu }^a\right)d{\bf k}\right],
\end{gather*}
leads to
\begin{multline*}
Z_{A}\propto \exp\left[ \dfr 12\int J_{\mu }^ak^{-2}J_{\mu }^a d{\bf k}\right]= \\ \exp\left[ \dfr 12\int J_{\mu }^a({\bf r}){|{\bf r-r'}|}^{-1}J_{\mu }^a({\bf r'}) d{\bf r}d{\bf r'}\right].
\end{multline*}
As a result, the system action can be reduced to the following form:
\begin{gather*}
S=S_{\phi }+S_J,
\end{gather*}
where $S_{\phi }$ is the scalar field part of the action, and
\begin{gather*}
S_J\approx C_1\sum\limits_{i=j} J_{\mu }^a({\bf r}_i)J_{\mu }^a({\bf r}_j)+C_2\sum\limits_{i\neq j}\dfr{J_{\mu }^a({\bf r}_i)J_{\mu }^a({\bf r}_j)}{|{\bf r}_i-{\bf r'}_j|},
\end{gather*}
where the disclination element $J_{\mu }^a({\bf r}_i)$ occupying the site $i$ at ${\bf r}_i$ position, is the contribution of the geometrical frustration \cite{Tarjus}.
The first term is the disclination core contribution to the action, and the second term is the contribution of the vortexes (disclinations) interaction \cite{Riv2}. According to the Kivelson--Tarjus theory the competition of these terms determines the properties of glass-forming liquids close to the glass transition.

\section{Conclusions}

According to the above theory one can offer the following physical picture of the glass transition processes:
when the temperature  approximates $T_g$, the correlation length of the gauge field diverges, while the correlation length of the order parameter is finite. The infinite correlation of the gauge field means that the relative rotations are correlated at an infinitely large distance. Then the growth of the order parameter field fluctuations becomes impossible, since the spins can not already turn independently of each other.

The statement about that the correlation length of the gauge field grows faster than the correlation length of the order parameter can be explained in the following way: without the frustrations the correlation lengths of both the gauge field and the order parameter field grow equally, because they are components of the same field. In the dual representation the system contains thermally activated vortexes, whose concentration tends to zero when $T \to T_c^+$. The priority growth rate of the correlation length of the gauge field becomes possible when the system contains frustration. The frustration induces vortexes additional to the thermally activated vortexes, as a result in the equilibrium state the vortex concentration is fixed non-zero. The reason for this is that the fluctuations of the gauge field are sure to develop around these sources. In the case of a nonzero concentration of these static sources the fluctuations around them can interflow, which leads to the faster growth of the effective correlation length of the gauge field. With the given source concentration and the appropriate correlation length of the order parameter (or the gauge field without frustrations) it leads to the formation of the percolation cluster which is associated with these sources. Thus, the effective correlation length of the gauge field becomes infinitely large, and the relative rotations at infinitely remote points become correlated and the relaxation time of the system infinitely grows up, but the order parameter correlation length remains to be relatively small. Therefore, the system freezes in a disordered state.

In conclusion, note that the above theoretical approach describes all general properties of glass transition very well: the expression for the temperature dependence of the heat capacity near the glass transition is in good qualitative agreement with the experimental data. The derived temperature dependences of the linear and non-linear susceptibilities behave quite properly at $T_g$, and the critical exponent of the non-linear susceptibilities is $\gamma \simeq 2$, that is a good estimation.  To crown it all, this theory reproduces the characteristic form of the $\langle \phi \phi\rangle_t$ correlation function dependence on time, and predicts the Vogel-Fulcher-Tammann law (\ref{VFT}) for the temperature dependence of the relaxation time.
The theory allows to answer some important questions that guide our presentation of the glass-transition theories \cite{Biroli}: first of all, according to the suggested  theory, the time of relaxation increases when $T\to T_g^{+}$ with non-Arrhenius law (\ref{VFT}) because of the features of the non-equilibrium dynamics of the frustrated system in the critical fluctuation region. The broad relaxation spectra are the result of the dynamic scaling in the fluctuation region, which results in the hierarchy of relaxation times. It is very difficult to say something about the relation between the kinetics and thermodynamics discussing the glass transition within the presented theory. One can only assume that $T_c$ can be considered as the Kauzmann temperature, $T_c=T_K$, because both of them are virtual parameters with similar meaning. In $T_c$ the non-frustrated system tends to the hypothetical ordered state with the minimum of the configuration entropy, but this state, as well as the Kauzmann state in $T=T_K$, is not physical. So the presented theory is based on the suggestion, that in the system there is an inaccessible (virtual) transition, $T_c$, but this is not an ``ideal glass transition'', since this inaccessible phase is assumed to be ordered. The collective motion in the theory can be represented as a vortexes motion, therefore the glass transition, when this motion becomes correlated and critically slows, is a collective phenomenon. The term ``order parameter'' has somewhat different meaning in comparison with the phase transition theory. Here the ``order parameter'' is the material field, which corresponds to the local ordering with the structure of the hypothetical ordered state as the local magnetization in spin systems. It is the order parameter of the inaccessible (virtual) phase transition, $T_c$. According to the suggested theory the dynamic correlation is directly associated with the gauge field correlation, because it does determine the divergence of the relaxation time of the order parameter. This theory gives a good geometrical representation of the glass transition as freezing the vortices network. It was developed for the spin systems, but it can also be reformulated for the description of the supercooled liquid--glass transition. In both cases it has the same topological nature, therefore, one can expect qualitatively identical results. In this sense the theory has a general character.

This work was partly supported by the RFBR grants No. 10-02-00882-a and No. 10-02-00700-a.

\end{document}